\documentstyle[12pt,epsf]{article}

\def\m{\mu}
\def\n{\nu}
\def\r{\rho}
\def\s{\sigma}
\def\p{\phi}
\def\vp{\varphi}
\def\th{\theta}
\def\b{\beta}
\def\a{\alpha}
\def\l{\lambda}
\def\pa{\partial}
\def\ep{\epsilon}

\def\ll{\left}
\def\rr{\right}
\def\fr{\frac}

\def\ci{\cite}
\def\bi{\bibitem}

\newcommand{\be}{\begin{eqnarray}}
\newcommand{\ee}{\end{eqnarray}}

\def\nn{\nonumber}
\def\sect{\section}

\begin{document}

\title
{\bf \LARGE Gravitating dyons and dyonic black holes 
in Einstein-Born-Infeld-Higgs model}

\author{Prasanta K. Tripathy\thanks{email: prasanta@iopb.res.in}\\
{\small\sl Institute of Physics, Bhubaneswar 751 005, India}}
\maketitle

\begin{abstract}

We find static spherically symmetric dyons in Einstein-Born-Infeld-Higgs
model in $3+1$ dimensions. The solutions share many features with the 
gravitating monopoles in the same model. In particular, they exist only 
up to some critical value of a parameter $\a $ related to the strength
of the gravitational interaction. We also study dyonic non-Abelian 
black holes. We analyse these solutions numerically.

\end{abstract} 

\sect{Introduction}
Gravitating non-Abelian solitons has attracted much attention recently
after the discovery of particle like solutions in Einstein-Yang-Mills (EYM) 
theory\ci{btnik} by Bartnik and McKinnon (for a review see ref.\ci{volkov}). 
Similar to the Bartnik-McKinnon particles, the EYM model also admit 
black hole solutions with nontrivial Yang-Mills connection \ci{vol,kun,bhl}. 
These are called non-Abelian black holes. Unfortunately, in spherically 
symmetric $SU(2)$ EYM theory there is neither globally regular solution 
nor black hole \ci{popp} with non-Abelian monopole or dyon charge other 
than the embedded Reissner-Nordstrom (RN) solution. However, monopole 
solutions were studied in the Einstein-Yang-Mills-Higgs (EYMH) model
\ci{peter,ortiz,dieter,lue} as a generalization of the 't~Hooft-Polyakov 
monopole\ci{toft,plkv} to see the effect of gravity on it. 
Numerical analysis showed that they exist only up to some critical value 
of a dimensionless parameter $\a $, characterising the strength of the 
gravitational interaction. The existence of these solutions were proved 
analytically\ci{dieter} for the case of infinite Higgs mass. Magnetically 
charged non-Abelian black holes were also found in this model\ci{lee,biz,sdm}. 
As a generalization of these above solutions, it was shown that the 
EYMH model admits dyons as well as dyonic black hole solutions\ci{tell,tel}. 
Like the monopole solutions these dyons also exist only up to some critical 
value of $\a$. The dependence on entropy of the non-Abelian balck hole 
solutions in various models has been studied\ci{tori,meda,tach}.

Recently the Born-Infeld\ci{born,infeld}  model has been widely studied in 
string theory. It was shown that the low energy effective action for 
D-brane\ci{dai} is described by the Born-Infeld action\ci{tseyt,stln,leigh}. 
Solitons of the Born-Infeld action correspond to the intersection of 
branes\ci{mald,gibb,west}. 
The Born-Infeld action is generalized for non-Abelian vector fields
and it was shown that the action is linearised \ci{brech} by the BPS 
configuration if it has a symmetrised trace structure\ci{nabi}.
The supersymmetric action for this was constructed and the connection
of BPS relation with supersymmetry was also discussed\ci{schapo}.
Some non BPS solutions\ci{gran,more} in different models containing 
Born-Infeld term were also studied.
The existence of gravitating monopoles as well as magnetically 
charged non-Abelian black holes is shown in Einstein-Born-Infeld-Higgs
(EBIH) model for static spherically symmetric fields\ci{tuku}.

Just as in EYMH case\ci{tell,tel}, both monopole and dyon solution exist 
in the same model, it is worthwhile enquiring if the EBIH model\ci{tuku}, 
which admit monopole and magnetically charged black hole also admit dyon 
and dyonic black hole. The purpose of this paper is to answer the question 
in affirmative. Motivated by these, we consider the EBIH model\ci{tuku} 
and study the dyon and dyonic non-Abelian black hole solutions. We find
that the solutions are similar to those for the EYMH case\ci{tell,tel}. 
In Sec.II we consider the model and find the equations of motion. 
In Sec.III we analyse them numerically and finally we conclude the results 
in Sec.IV.

\sect{The Model}

We consider the following Einstein-Born-Infeld-Higgs action\ci{tuku} for 
$SU(2)$ fields with the Higgs field in the adjoint representation

\be
S = \int d^4x \sqrt{-g} \ll[L_G + L_{BI} + L_H \rr]
\ee
with
\be
L_G & = & \fr{1}{16\pi G}{\cal R} , \nn \\
L_H & = & -\fr{1}{2} D_{\m}\p ^a D^{\m}\p ^a
          -\fr{e^2g^2}{4}\ll(\p ^a\p ^a - v^2 \rr)^2 \nn
\ee
and the non-Abelian Born-Infeld Lagrangian,
\be
L_{BI} = \b ^2 Str\ll( 1 - \sqrt{ 1
+ \fr{1}{2\b ^2}F_{\m\n}F^{\m\n}
- \fr{1}{8\b ^4}\ll(F_{\m\n}\tilde{F}^{\m\n}\rr)^2}\rr) \nn
\ee
where
\be
D_{\m}\p ^a = \pa _{\m}\p ^a + e \ep ^{abc} A_{\m}^b\p ^c , \nn
\ee
\be
F_{\m\n} = F_{\m\n}^a t^a
= \ll(\pa _{\m}A_{\n}^a - \pa_{\n}A_{\m}^a
+ e \ep ^{abc}A_{\m}^bA_{\n}^c\rr)t^a \nn
\ee
and 
\be
\tilde{F}_{\m\n} = \fr{1}{2}\sqrt{-g}~~\epsilon _{\m\n\r\s}F^{\r\s}
\ee
The symmetric trace is defined as

\be
Str\ll(t_1,t_2...,t_n\rr) = \fr{1}{n!}
\sum tr\ll(t_{i_1}t_{i_2}...t_{i_n}\rr) \nn .
\ee
where the sum is over all permutations on the product of the $n$ 
generators $t_i$. Unlike the monopole case we have here 
$F_{\m\n}\tilde{F}^{\m\n} \ne 0 $.  Expanding the square root in 
powers of $\fr{1}{\b ^2}$ and keeping up to order  $\fr{1}{\b ^2}$ 
we have the Born-Infeld Lagrangian

\be
L_{BI} = -\fr{1}{4}F_{\m\n}^a F^{a \m\n}
+\fr{1}{96\b ^2}\ll[\ll(F_{\m\n}^a F^{a \m\n}\rr)^2
+2F_{\m\n}^a F_{\r\s}^a F^{b \m\n}F^{b \r\s}
\right. \nn \\ \left.
+\ll(F_{\m\n}^a \tilde{F}^{a \m\n}\rr)^2
+2F_{\m\n}^a F_{\r\s}^a \tilde{F}^{b \m\n}\tilde{F}^{b \r\s}\rr]
+ O(\fr{1}{\b ^4}) .
\ee
We consider the static spherical symmetric solutions for which the metric 
can be parametrized as\ci{komar,dieter} 

\be
ds^2 = -e ^{2\n(R)}dt^2 + e ^{2\l(R)}dR^2
+ r^2(R)(d\th ^2 + \sin ^2\th d\vp ^2)
\ee
The ansatz for the gauge and the scalar fields are 

\be
&& A_{t}^a(R) = e_{R}^a v J(R), ~~ A_{R}^a = 0, \nn \\
&& A_{\th}^a = e_{\vp}^a\fr{W(R) - 1}{e}, ~~ 
A_{\vp}^a = -e_{\th}^a\fr{W(R) - 1}{e}\sin\th ,
\ee
and
\be
\p ^a = e_{R}^a v H(R) .
\ee
We obtain the following expression for the Lagrangian after putting
the above ansatz in Eq.(1) and rescaling $ R \rightarrow R/ev, 
\b \rightarrow \b ev^2 $ and $ r(R) \rightarrow r(R)/ev $,

\be
\int dR ~~ e^{\n +\l}\ll[\fr{1}{2}\ll(1
+ e^{-2\l}\ll((r')^2 + \n '(r^2)'\rr)\rr)
\right. \nn \\ \left.
- \a ^2 \ll(e^{-2\l} ( V_1 - U_1 ) 
- e^{-4\l} V_2 - U_2 + V_3 - U_3 \rr)\rr],
\ee
where we define $\a ^2 = 4\pi Gv^2$, and 
\be
V_1 = (W')^2 +\fr{1}{2}r^2(H')^2 
-\fr{1}{6\b ^2r^4} (W')^2 (W^2 - 1)^2,
\ee

\be
V_2 = \fr{1}{3\b ^2r^2} (W')^4
\ee

\be
V_3 = \fr{1}{2 r^2} (W^2 - 1)^2 
+ W^2H^2 
+ \fr{g^2r^2}{4}(H^2 - 1)^2 
- \fr{1}{8\b ^2 r^6} (W^2 - 1)^4 
\ee

\be
U_1 = &&\fr{1}{4\b ^2 r^2} e^{-2\n} (J')^2 (W^2 - 1)^2
+\fr{1}{6\b ^2} e^{-4\n} (J J' W)^2 \nn \\
&& +\fr{2}{3\b ^2 r^2} e^{-2\n} J J' W W' (W^2 - 1)
+ \fr{2}{3\b ^2 r^2} e^{-2\n} (J W W')^2 ,
\ee

\be
U_2 = && \fr{1}{2} e^{-2(\l +\n )} r^2 (J')^2 
- \fr{1}{6\b ^2} e^{-2\n -4\l} (J'W')^2 \nn \\
&& + e^{-2\n} (JW)^2 
- \fr{1}{6\b ^2 r^4} e^{-2\n} (JW)^2 (W^2 - 1)^2
\ee
and
\be
U_3 = \fr{1}{8\b ^2} e^{-4(\n +\l )} r^2 (J')^4
+\fr{1}{3\b ^2 r^2} e^{-4\n} (JW)^4
\ee
Here the prime denotes differentiation with respect to $R$.
The dimensionless parameter $\a $ can be expressed as the mass ratio
\be
\a = \sqrt{4\pi}\fr{M_W}{eM_{Pl}}
\ee
with the gauge field mass $M_W = e v $ and the Planck mass
$M_{Pl} = 1/ \sqrt{G} $ . Note that the Higgs mass $M_H = \sqrt{2} g e v $.
In the limit of $\b \rightarrow \infty $ the above action reduces to
that of the EYMH model. For the case of $\a = 0$ we 
must have $\n (R) = 0 = \l (R)$ which corresponds to the flat space 
Born-Infeld-Higgs theory.
From now on we restrict ourselves to the gauge choice $r(R) = R $, 
corresponding to the Schwarzschild-like coordinates and rename $R = r $.
We define $A = e^{\n + \l}$ and $N = e^{-2\l}$. Integrating the $tt$ 
component of the energy-momentum tensor, which is obtained by varying the 
matter field action with respect to the metric, we get the mass of the 
dyon equal to $M/evG$ where
\be
M = \a ^2 \int_{0}^{\infty} dr \ll(N V_1 
- N^2 V_2 + U_2 + V_3 +3 U_3 - U_4 \rr)
\ee
with 
\be
U_4 = && \fr{15}{4\b ^2 r^2} \ll(\fr{J'}{A}\rr)^2 (W^2 - 1)^2
-\fr{1}{2\b ^2 N} \ll(\fr{JJ'}{A^2} W\rr)^2 \nn \\
&& +\fr{14}{3\b ^2 r^2} \fr{JJ'}{A^2} W W' (W^2 - 1)
+ \fr{10}{\b ^2 r^2} \ll(\fr{J}{A} W W'\rr)^2 
\ee
The Abelian field strength ${\cal F}_{\m\n}$, as pointed out by 't Hooft,
can be defined as 
\be
{\cal F}_{\m\n} = \fr{\p ^aF_{\m\n}^a}{\mid \p \mid}
- \fr{1}{e\mid\p\mid ^3}\epsilon^{abc}\p ^aD_{\m}\p ^bD_{\n}\p ^c. \nn
\ee
One can show that in our case the magnetic field 
\be B^i = \fr{1}{2}\epsilon ^{ijk}{\cal F}_{jk} \nn \ee is equal to
$ {e_{r}^{i}}/{er^2} $ with a total flux
$4\pi /e $ and unit magnetic charge, while
the electric field $E^i = - {\cal F}_{0}^{~i} $ is equal to 
${e}_{r}^{i}(J/er)'$ with electric charge $Q = \int ds_i{\cal F}_{0}^{~i}$.

The $tt$ and $rr$ components of the Einstein's equations are 
\be
&& \fr{1}{2}[1 - (rN)'] 
= \a ^2\ll( N V_1 - N^2 V_2 + U_2 + V_3 + 3 U_3 - U_4 \rr) \\
&& \fr{A'}{A} = \fr{2\a ^2}{r}\ll(V_1 - 2 N V_2 + U_5 \rr)
\ee
with 
\be
U_5 & = & \fr{1}{6\b ^2}\ll[
\ll(\fr{JJ'W}{NA^2}\rr)^2 + \ll(\fr{J'W'}{A}\rr)^2\rr]
+\fr{2}{3\b ^2 r^2}N\ll(\fr{JW}{NA}\rr)^4 \nn \\
& & +\ll(\fr{JW}{NA}\rr)^2\ll(1 - \fr{1}{6\b ^2 r^4} (W^2 - 1)^2\rr)
\ee
and the matter field equations are 

\be
U_6' = &&\fr{1}{3\b ^2 N}\ll(\fr{J'}{A}\rr)^2\fr{JW^2}{A}
+\fr{2}{3\b ^2 r^2}\fr{J'}{A}W W' (W^2 - 1)
+2\fr{JW^2}{AN} \nn \\
&& +\fr{4}{3\b ^2 r^2 N^2}\ll(\fr{J'}{A}\rr)^3 W^4 
+\fr{4}{3\b ^2 r^2}\fr{J}{A}(W W')^2 
-\fr{1}{3\b ^2 r^4 N}\fr{J}{A}W^2 (W^2 - 1)^2 
\ee

\be
\fr{1}{A}(A V_4)' = && W \ll(\fr{J'}{A}\rr)^2
\ll(\fr{1}{\b ^2r^2} (W^2 - 1) 
+\fr{1}{3\b ^2 N} \ll(\fr{J}{A}\rr)^2\rr) 
+\fr{2}{3\b ^2 r^2} \fr{JJ'}{A^2}W'(3W^2 - 1) \nn \\
&& + \fr{4}{3\b ^2r^2}\ll(\fr{J}{A}\rr)^4\fr{W^3}{N^2}
+\ll(\fr{J}{A}\rr)^2\fr{W}{N}\ll( 2 
+ \fr{4N}{3\b ^2 r^2}(W')^2 
-\fr{1}{3\b^2r^4}(W^2 - 1)^2 \rr. \nn \\
&& \ll.  -\fr{2}{3\b ^2 r^2} W^2(W^2 - 1)\rr) 
-W \ll(
\fr{2}{r^2}(W^2 - 1) + 2 H^2 - \fr{(W^2 - 1)^3}{\b ^2 r^6}
\rr. \nn \\ && \ll.
-\fr{2N(W')^2}{3\b ^2 r^4} (W^2 - 1) \rr) 
\ee

\be
(ANr^2H')'  = A H \ll(2W^2 + g^2r^2(H^2 - 1)\rr)
\ee
with
\be
U_6 = && r^2\fr{J'}{A} + \fr{1}{2\b ^2}r^2\ll(\fr{J'}{A}\rr)^3
+\fr{1}{2\b ^2r^2}\fr{J'}{A}(W^2 - 1)^2 \nn \\
&& +\fr{1}{3\b ^2 N} \fr{J'}{A}\ll(\fr{JW}{A}\rr)^2 
-\fr{1}{3\b ^2}N\fr{J'}{A} (W')^2 
+\fr{2}{3\b ^2 r^2}\fr{J}{A} W W' (W^2 - 1)
\ee
and
\be
V_4 = && -\fr{1}{3\b ^2} \ll(\fr{J'}{A}\rr)^2 N W'
+\fr{2}{3\b ^2 r^2} \fr{JJ'}{A^2} W (W^2 - 1)
+\fr{4}{3\b ^2r^2}\ll(\fr{JW}{A}\rr)^2 W' \nn \\
&& - 2NW' + \fr{1}{3\b ^2 r^4} N W' (W^2 - 1)^2
+ \fr{4}{3\b ^2 r^2} N^2 (W')^3
\ee
As expected, the above equations reduce to those for the monopoles\ci{tuku}
if we put $J = 0$. Also they reduce to the corresponding dyonic equations for 
EYMH case\ci{tell,tel} in the limit of $\b \rightarrow \infty $. The equations
for EYMH system admit embedded RN solutions. In this case we have the following
generalized embedded RN solutions satisfying the above equations of motion for 
finite values of $\b $ :

\be
&& W(r) = 0, ~~~ H(r) = 1, ~~~ A(r) = 1, \nn \\
&& N(r) = 1 - \fr{2M}{r} 
+\a ^2\ll( \fr{1}{r^2}(1 + Q^2)
-\fr{1}{20\b ^2 r^6}(1 + 34Q^2 + Q^4)\rr) 
+O\ll(\fr{1}{\b ^4}\rr) \nn \\
&& J(r) = J_{\infty} - \fr{Q}{r} 
+ \fr{Q}{10\b ^2r^5}(1 + Q^2)
+O\ll(\fr{1}{\b ^4}\rr)
\ee
The mass is related to the horizon radius $r_h$ by the following expression
\be
M = \fr{\a ^2}{2}\ll( r_h + \fr{1}{r_h}(1 + Q^2)
-\fr{1}{20\b ^2r^5}(1 + 34Q^2 + Q^4)\rr)
\ee
obtained by demanding $N(r_h) = 0$. Inverting this relation, we obtain
\be
r_h = r_0 + \fr{1+34Q^2+Q^4}{40\b ^2r_{0}^{4}(r_0-M)} 
+ O\ll(\fr{1}{\b ^4}\rr)
\ee
with $r_0 = M + \sqrt{M^2 - \a ^2(1+Q^2)}$ and hence the solution exist for
$ M > \a \sqrt{1 + Q^2}$. This implies that unlike the $\b = \infty $
case where the extremal black holes exist with $M = \a \sqrt{1 + Q^2} $,
the equality does not hole for finite $\b $.

\sect{Solutions}
\subsection{Dyons}
We first consider the globally regular solutions to the equations of 
motion. For the solutions to be regular at the origin we have
\be
N(0) = 1 , 
\ee
and 
\be
H(0) = 0, ~~~ W(0) = 1, ~~~ J(0) = 0.
\ee
For asymptotically flat solutions we require 
\be
N(\infty ) = 1
\ee
and then for finite energy configuration  the matter fields approach the 
values 
\be
H(\infty ) = 1, ~~~ W(\infty )  = 0, ~~~ J(\infty ) = J_{\infty}
\ee
where $J_{\infty}$ is a constant.

We solved the equations numerically for various values of $\a $ 
keeping $\b $ and $g $ fixed. For large $r $ these solutions converge
to their asymptotic values. For $\a = 0 $ they correspond to the flat 
space dyons\ci{silv}. The behaviour of the solutions for $\a \ge 0 $ are 
similar to those for the globally regular monopole solutions\ci{tuku}. 
They exist up to some critical value of $\a $ above which there is no 
solution. The minimum of the metric function $N$ is 
found to be decreasing as $\a $ is increased gradually from zero to 
$\a _{max}$ and becomes zero at $\a = \a _{max} $. For $g = .1$ and 
$\b = 4$ we find $\a _{max}\sim 1.7$. The flat space solutions for 
dyons, corresponding to the value $\a = 0$ are given in Figs.1 and 2.
The profile for the fields for various values of $\a, (0<\a <\a _{max})$ 
with $g = .1 $ and $\b = 4 $ are shown in Figs.3,4,5 and 6.

\subsection{Dyonic black holes}

Apart from the globally regular solutions the EBIH model also  admit 
dyonic black holes. The event horizon is charecterised by some finite
$r_h$ for which $N(r_h) = 0$ and $A(r_h)$ is finite. The matter functions
at $r_h$ must satisfy the following conditions.

\be
&& J(r_h) = 0,   \\
&&W'N'\ll.\ll( - 2 - \fr{1}{3\b ^2}\ll(\fr{J'}{A}\rr)^2
+\fr{1}{3\b ^2 r^4}(W^2 -1)^2\rr)\rr|_{r_h} \nn \\
&& = \ll.\fr{1}{r^2}W(W^2 - 1)
\ll( - 2 + \fr{1}{3\b ^2}\ll(\fr{J'}{A}\rr)^2
+\fr{1}{\b ^2 r^4}(W^2 -1)^2\rr) - 2H^2 W \rr|_{r_h}  \\
&& \ll. r^2H'N'\rr|_{r_h}
= \ll. H\ll( 2W^2 + g^2 r^2 (H^2 - 1)\rr)\rr|_{r_h}  \\
&&\ll.\ll(r - \fr{3}{2\b ^2}r^2 A'\rr)\ll(\fr{J'}{A}\rr)^2
+ 2r -\fr{1}{\b ^2}(W^2 - 1)^2\ll(\fr{1}{r^3} + \fr{1}{2r^2}A'\rr) 
- r^2 A' \rr. \nn \\ 
&& \ll. - \fr{1}{3\b ^2} N'(W')^2   
+\fr{2}{\b ^2 r^2} WW'(W^2 - 1) \rr|_{r_h}
= \ll. \fr{W^2}{N'}\ll(2 - \fr{1}{3\b ^2 r^4}(W^2 - 1)^2\rr)\rr|_{r_h} 
\ee

\be
\ll. 1 - rN'\rr|_{r_h}
&=& 2\a^2\ll.\ll(
\ll(\fr{J'}{A}\rr)^2\ll(\fr{1}{2}r^2 
- \fr{15}{4\b ^2r^2}(W^2 - 1)^2\rr)
+\fr{3}{8\b ^2} r^2\ll(\fr{J'}{A}\rr)^4 
\rr.\rr. \nn \\
&&+ \fr{1}{2r^2}(W^2 - 1)^2 
\ll.\ll. -\fr{1}{8\b ^2r^6}(W^2 - 1)^4
+ H^2W^2 + \fr{1}{4} g^2r^2(H^2 - 1)^2\rr)\rr|_{r_h} ~~
\ee
and

\be
& & rA'|_{r_h}
= 2\a ^2\ll.\ll(
\ll(\fr{J'}{A}\rr)^2\ll(\fr{1}{6\b ^2}(W')^2 
+ \fr{W^2}{(N')^2}\ll( 1 
- \fr{1}{6\b ^2 r^4}(W^2 - 1)^2\rr)\rr)
\rr.\rr. \nn \\ & & \ll.\ll.
+\fr{1}{6\b ^2}\ll(\fr{W}{N'}\rr)^2\ll(\fr{J'}{A}\rr)^4
+ (W')^2 + \fr{1}{2}r^2(H')^2
-\fr{1}{6\b ^2r^4}(W')^2(W^2 - 1) \rr)\rr|_{r_h}
\ee
They follow the same behaviour as the globally regular solutions in the 
region of large $r$ as given in Eqs.(30) and (31). With these boundary 
conditions we solve the equations numerically. 
They have many features in common with the magnetically 
charged black holes. In particular, for $r_h$ close to zero, the solutions 
approach to the regular dyon solutions. The profile for the fields are 
given in Fig.7. 

\sect{Conclusion}

In the present work we have investigated gravitating dyons in the EBIH model.
We derived a generalized expression for the embedded RN solitons. Apart from
this embedded Abelian solution there are also non-Abelian solutions. In
particular, we found that globally regular dyons exist only up to some critical 
value $\a _{max}$ of the parameter $\a $. We also found the dyonic non-Abelian 
black hole solutions. The solutions are similar to the corresponding monopoles
and magnetically charged non-Abelian black holes. It would be interesting 
to prove the existence of the solutions analytically. 

\sect{Acknowledgements}

I am indebted to Avinash Khare for many helpful discussions as well as for 
a critical manuscript reading.

\newpage

\begin{figure}
\vspace{-1in}
\vglue.1in
\makebox{
\epsfxsize=9in
\epsfbox{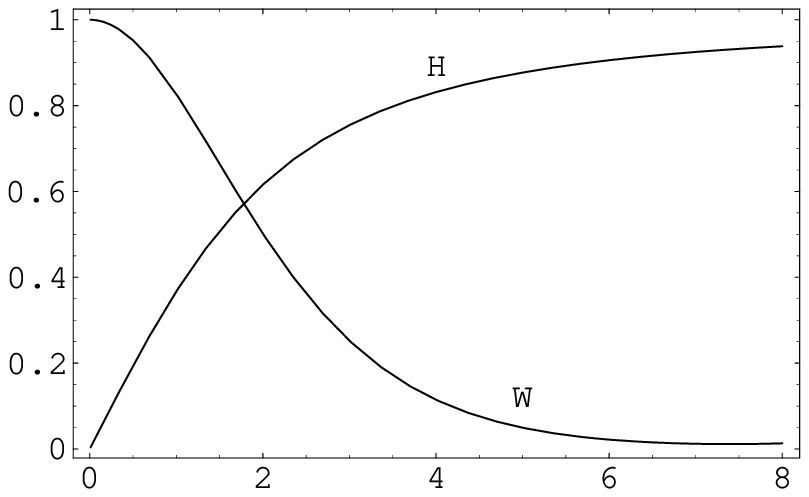}
}
\vspace{-9in}
\caption{
Flat space dyon solution for $\b = 4$ and $g = .1$ 
(Variation of Higgs field $H$ and gauge field $W$ with $r$). 
}


\vglue.1in
\makebox{
\epsfxsize=9in
\epsfbox{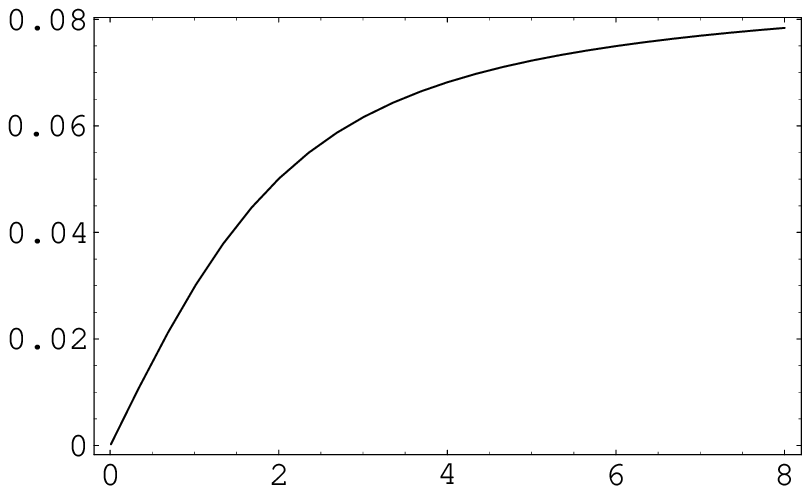}
}
\vspace{-9in}
\caption{
Variation of the field $J$ as a function of $r$ for the above 
dyon solution in flat space.
}

\end{figure}

\newpage

\begin{figure}
\vspace{-.5cm}
\vglue.1in
\makebox{
\epsfxsize=9in
\epsfbox{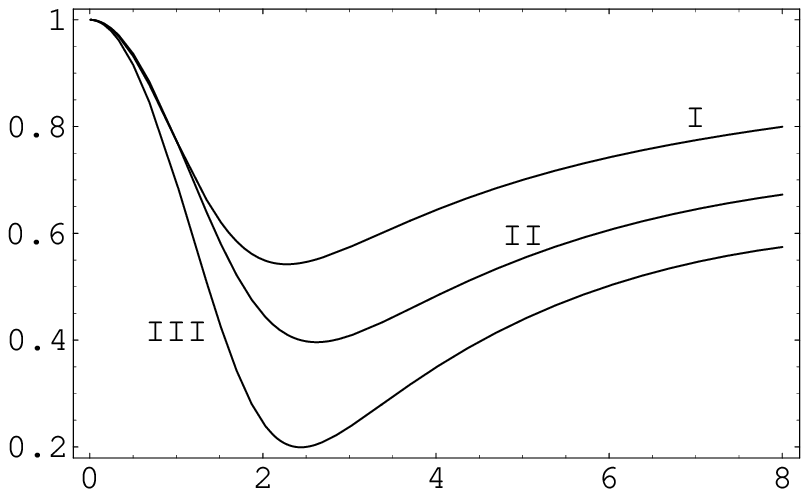}
}
\vspace{-9in}
\caption{
Plot for the metric function $N$ as a function of $r$ for 
$\b = 4$ and $g = .1 $ for various values of $\a $.  
Curve I is for $\a = 1$, curve II for $\a = 1.38 $ 
and curve III for $\a = 1.64 $.
}


\vglue.1in
\makebox{
\epsfxsize=9in
\epsfbox{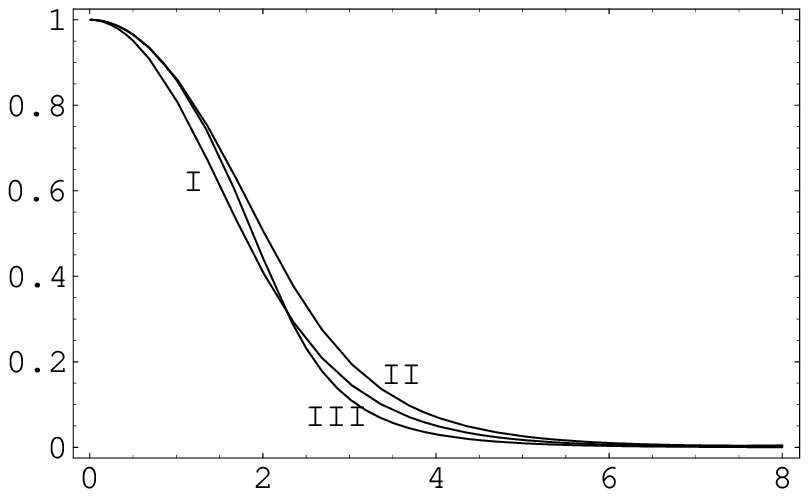}
}
\vspace{-9in}
\caption{
Plot for the gauge field $W$ as a function of $r$ for 
$\b = 4$ and $g = .1 $ for various values of $\a $.  
Curve I is for $\a = 1$, curve II for $\a = 1.38 $ 
and curve III for $\a = 1.64 $.
}
\end{figure}

\newpage

\begin{figure}
\vspace{-1in}
\vglue.1in
\makebox{
\epsfxsize=9in
\epsfbox{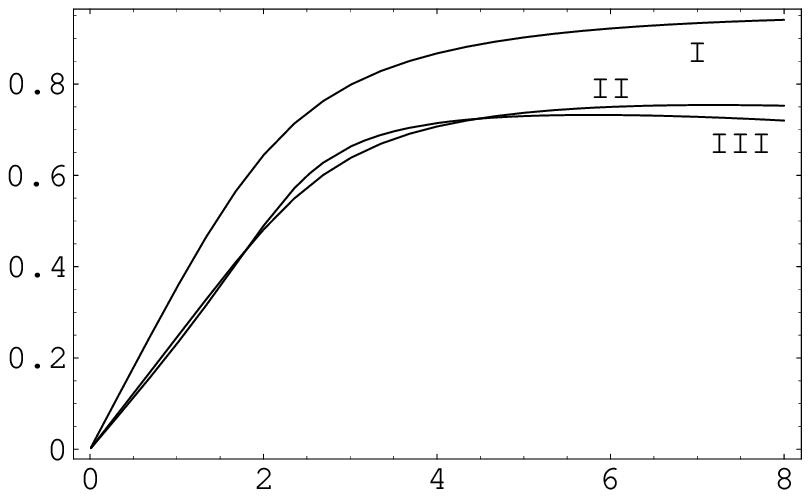}
}
\vspace{-9in}
\caption{
Plot for the Higgs field $H$ as a function of $r$ for 
$\b = 4$ and $g = .1 $ for various values of $\a $.  
Curve I is for $\a = 1$, curve II for $\a = 1.38 $ 
and curve III for $\a = 1.64 $.
}


\vglue.1in
\makebox{
\epsfxsize=9in
\epsfbox{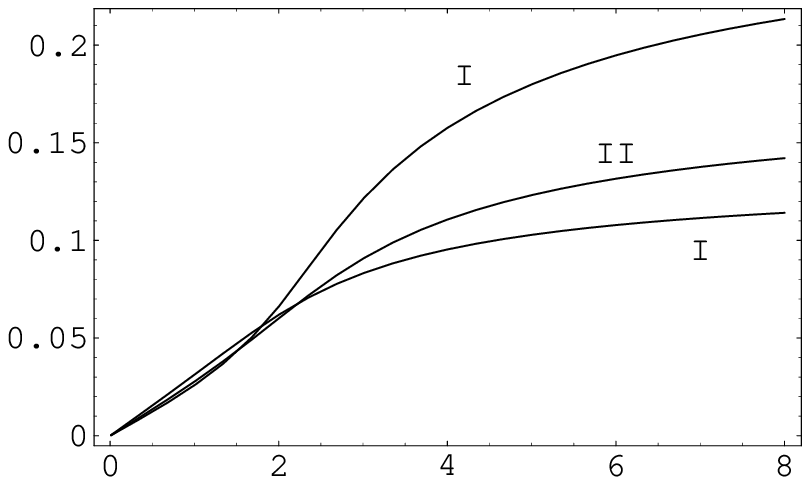}
}
\vspace{-9in}
\caption{
Plot for the field $J$ as a function of $r$ for 
$\b = 4$ and $g = .1 $ for various values of $\a $.  
Curve I is for $\a = 1$, curve II for $\a = 1.38 $ 
and curve III for $\a = 1.64 $.
}

\end{figure}
\newpage

\begin{figure}
\vspace{-1in}
\vglue.1in
\makebox{
\epsfxsize=9in
\epsfbox{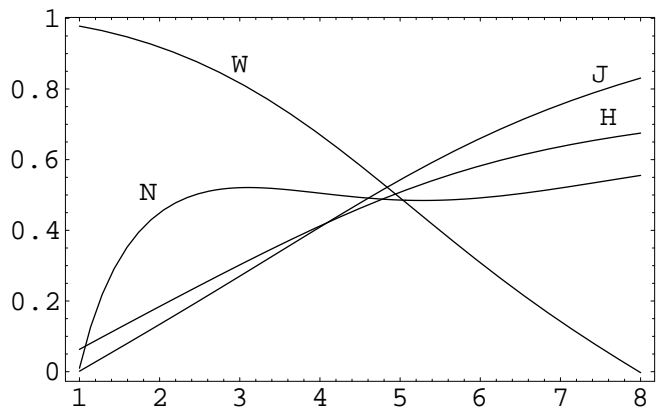}
}
\vspace{-9in}
\caption{
Black hole solution for $g = 0,~~ \a = 1$ and $ \b = 4$ with 
the horizon radius $r_h = .99 $.
}
\end{figure}

\end{document}